\PassOptionsToPackage{unicode}{hyperref}
\PassOptionsToPackage{hyphens}{url}
\PassOptionsToPackage{dvipsnames,svgnames,x11names}{xcolor}
\documentclass[
  12pt,
]{interact}
\usepackage{xcolor}
\usepackage{amsmath,amssymb}
\usepackage{iftex}
\ifPDFTeX
  \usepackage[T1]{fontenc}
  \usepackage[utf8]{inputenc}
  \usepackage{textcomp} 
\else 
  \usepackage{unicode-math} 
  \defaultfontfeatures{Scale=MatchLowercase}
  \defaultfontfeatures[\rmfamily]{Ligatures=TeX,Scale=1}
\fi
\usepackage{lmodern}
\ifPDFTeX\else
\fi
\IfFileExists{upquote.sty}{\usepackage{upquote}}{}
\IfFileExists{microtype.sty}{
  \usepackage[]{microtype}
  \UseMicrotypeSet[protrusion]{basicmath} 
}{}
\usepackage{setspace}
\makeatletter
\@ifundefined{KOMAClassName}{
  \IfFileExists{parskip.sty}{%
    \usepackage{parskip}
  }{
    \setlength{\parindent}{0pt}
    \setlength{\parskip}{6pt plus 2pt minus 1pt}}
}{
  \KOMAoptions{parskip=half}}
\makeatother
\makeatletter
\ifx\paragraph\undefined\else
  \let\oldparagraph\paragraph
  \renewcommand{\paragraph}{
    \@ifstar
      \xxxParagraphStar
      \xxxParagraphNoStar
  }
  \newcommand{\xxxParagraphStar}[1]{\oldparagraph*{#1}\mbox{}}
  \newcommand{\xxxParagraphNoStar}[1]{\oldparagraph{#1}\mbox{}}
\fi
\ifx\subparagraph\undefined\else
  \let\oldsubparagraph\subparagraph
  \renewcommand{\subparagraph}{
    \@ifstar
      \xxxSubParagraphStar
      \xxxSubParagraphNoStar
  }
  \newcommand{\xxxSubParagraphStar}[1]{\oldsubparagraph*{#1}\mbox{}}
  \newcommand{\xxxSubParagraphNoStar}[1]{\oldsubparagraph{#1}\mbox{}}
\fi
\makeatother

\usepackage{longtable,booktabs,array}
\usepackage{calc} 
\usepackage{etoolbox}
\makeatletter
\patchcmd\longtable{\par}{\if@noskipsec\mbox{}\fi\par}{}{}
\makeatother
\IfFileExists{footnotehyper.sty}{\usepackage{footnotehyper}}{\usepackage{footnote}}
\makesavenoteenv{longtable}
\usepackage{graphicx}
\makeatletter
\newsavebox\pandoc@box
\newcommand*\pandocbounded[1]{
  \sbox\pandoc@box{#1}%
  \Gscale@div\@tempa{\textheight}{\dimexpr\ht\pandoc@box+\dp\pandoc@box\relax}%
  \Gscale@div\@tempb{\linewidth}{\wd\pandoc@box}%
  \ifdim\@tempb\p@<\@tempa\p@\let\@tempa\@tempb\fi
  \ifdim\@tempa\p@<\p@\scalebox{\@tempa}{\usebox\pandoc@box}%
  \else\usebox{\pandoc@box}%
  \fi%
}
\def\fps@figure{htbp}
\makeatother

\NewDocumentCommand\citeproctext{}{}

\makeatletter
 \let\@cite@ofmt\@firstofone
 \def\@biblabel#1{}
 \def\@cite#1#2{{#1\if@tempswa , #2\fi}}
\makeatother
\newlength{\cslhangindent}
\newlength{\csllabelwidth}
\newenvironment{CSLReferences}[2] 
 {\begin{list}{}{%
  \setlength{\itemindent}{0pt}
  \setlength{\leftmargin}{0pt}
  \setlength{\parsep}{0pt}
  \ifodd #1
   \setlength{\leftmargin}{\cslhangindent}
   \setlength{\itemindent}{-1\cslhangindent}
  \fi
  \setlength{\itemsep}{#2\baselineskip}}}
 {\end{list}}
\usepackage{calc}

\providecommand{\tightlist}{%
  \setlength{\itemsep}{0pt}\setlength{\parskip}{0pt}}

\usepackage{booktabs}
\usepackage{longtable}
\usepackage{array}
\usepackage{multirow}
\usepackage{wrapfig}
\usepackage{float}
\usepackage{colortbl}
\usepackage{pdflscape}
\usepackage{tabu}
\usepackage{threeparttable}
\usepackage{threeparttablex}
\usepackage[normalem]{ulem}
\usepackage{makecell}
\usepackage{xcolor}
\usepackage{caption}
\usepackage{anyfontsize}
\usepackage{orcidlink}
\usepackage{algorithm}
\usepackage{float}
\usepackage{amsmath}
\usepackage{amsthm}
\usepackage{amssymb}
\theoremstyle{plain}
\newtheorem{defn}{\protect\definitionname}

\providecommand{\definitionname}{Definition}
\providecommand{\propositionname}{Proposition}
\makeatletter
\@ifpackageloaded{tcolorbox}{}{\usepackage[skins,breakable]{tcolorbox}}
\@ifpackageloaded{fontawesome5}{}{\usepackage{fontawesome5}}
\definecolor{quarto-callout-color}{HTML}{909090}
\definecolor{quarto-callout-note-color}{HTML}{0758E5}
\definecolor{quarto-callout-important-color}{HTML}{CC1914}
\definecolor{quarto-callout-warning-color}{HTML}{EB9113}
\definecolor{quarto-callout-tip-color}{HTML}{00A047}
\definecolor{quarto-callout-caution-color}{HTML}{FC5300}
\definecolor{quarto-callout-color-frame}{HTML}{acacac}
\definecolor{quarto-callout-note-color-frame}{HTML}{4582ec}
\definecolor{quarto-callout-important-color-frame}{HTML}{d9534f}
\definecolor{quarto-callout-warning-color-frame}{HTML}{f0ad4e}
\definecolor{quarto-callout-tip-color-frame}{HTML}{02b875}
\definecolor{quarto-callout-caution-color-frame}{HTML}{fd7e14}
\makeatother
\makeatletter
\@ifpackageloaded{caption}{}{\usepackage{caption}}
\AtBeginDocument{%
\ifdefined\contentsname
  \renewcommand*\contentsname{Table of contents}
\else
  \newcommand\contentsname{Table of contents}
\fi
\ifdefined\listfigurename
  \renewcommand*\listfigurename{List of Figures}
\else
  \newcommand\listfigurename{List of Figures}
\fi
\ifdefined\listtablename
  \renewcommand*\listtablename{List of Tables}
\else
  \newcommand\listtablename{List of Tables}
\fi
\ifdefined\figurename
  \renewcommand*\figurename{Figure}
\else
  \newcommand\figurename{Figure}
\fi
\ifdefined\tablename
  \renewcommand*\tablename{Table}
\else
  \newcommand\tablename{Table}
\fi
}
\@ifpackageloaded{float}{}{\usepackage{float}}
\floatstyle{ruled}
\@ifundefined{c@chapter}{\newfloat{codelisting}{h}{lop}}{\newfloat{codelisting}{h}{lop}[chapter]}
\floatname{codelisting}{Listing}

\makeatother
\makeatletter
\@ifpackageloaded{caption}{}{\usepackage{caption}}
\@ifpackageloaded{subcaption}{}{\usepackage{subcaption}}
\makeatother
\usepackage{bookmark}
\IfFileExists{xurl.sty}{\usepackage{xurl}}{} 
\hypersetup{
  pdftitle={Squintability and Other Metrics for Assessing Projection Pursuit Indexes, and Guiding Optimization Choices},
  pdfauthor={H. Sherry Zhang; Dianne Cook; Nicolas Langrené; Jessica Wai Yin Leung},
  pdfkeywords={Projection Pursuit Guided Tour (PPGT), Jellyfish Search
Optimizer (JSO)},
  colorlinks=true,
  linkcolor={blue},
  filecolor={Maroon},
  citecolor={Blue},
  urlcolor={Blue},
  pdfcreator={LaTeX via pandoc}}

\title{Squintability and Other Metrics for Assessing Projection Pursuit
Indexes, and Guiding Optimization Choices}
\author{H. Sherry Zhang$\textsuperscript{1}$, Dianne
Cook$\textsuperscript{2}$, Nicolas
Langrené$\textsuperscript{3}$, Jessica Wai Yin
Leung$\textsuperscript{2}$}

\thanks{CONTACT: H. Sherry
Zhang. Email: \href{mailto:huize.zhang@austin.utexas.edu}{\nolinkurl{huize.zhang@austin.utexas.edu}}. }
\begin{document}
\captionsetup{labelsep=space}
\maketitle
\textsuperscript{1} Department of Statistics and Data
Sciences, University of Texas at Austin, Austin, United
States\\ \textsuperscript{2} Department of Econometrics and Business
Statistics, Monash
University, Melbourne, Australia\\ \textsuperscript{3} Department of
Mathematical Sciences, Guangdong Provincial/Zhuhai Key Laboratory of
Interdisciplinary Research and Application for Data Science, Beijing
Normal-Hong Kong Baptist University, Zhuhai, China
\begin{abstract}
The projection pursuit (PP) guided tour optimizes a criterion function,
known as the PP index, to gradually reveal projections of interest from
high-dimensional data through animation. Optimization of some PP indexes
can be non-trivial, if they are non-smooth functions, or when the
optimum has a small ``squint angle'', detectable only from close
proximity. Here, measures for calculating the smoothness and
squintability properties of the PP index are defined. These are used to
investigate the performance of a recently introduced swarm-based
algorithm, Jellyfish Search Optimizer (JSO), for optimizing PP indexes.
The performance of JSO in detecting the target pattern (pipe shape) is
compared with existing optimizers in PP. Additionally, JSO's performance
on detecting the sine-wave shape is evaluated using different PP indexes
(hence different smoothness and squintability) across various data
dimensions (d = 4, 6, 8, 10, 12) and JSO hyper-parameters. We observe
empirically that higher squintability improves the success rate of the
PP index optimization, while smoothness has no significant effect. The
JSO algorithm has been implemented in the R package, \texttt{tourr}, and
functions to calculate smoothness and squintability measures are
implemented in the \texttt{ferrn} package.
\end{abstract}
\begin{keywords}
\def\sep{;\ }
Projection Pursuit Guided Tour (PPGT)\sep 
Jellyfish Search Optimizer (JSO)
\end{keywords}

\setstretch{2}
\section{Introduction}\label{introduction}

Projection pursuit (PP) (Kruskal 1969; Friedman and Tukey 1974; Huber
1985) is a dimension reduction technique aimed at identifying
informative linear projections of data. This is useful for exploring
high-dimensional data and creating plots of the data that reveal the
main features to use for publication. The method involves optimizing an
objective function known as the PP index (e.g., Hall 1989; Cook, Buja,
and Cabrera 1993; Lee and Cook 2010; Loperfido 2018, 2020), which
defines the criterion for what constitutes interesting or informative
projections. Let \(X \in \mathbb{R}^{n\times p}\) be the data matrix,
\(A \in\mathbb{R}^{p \times d}\) be an orthonormal matrix in the Stiefel
manifold \(\mathcal{A} = V_d(\mathbb{R}^p)\). The projection \(Y = XA\)
is a linear transformation of the \(p\)-dimensional data to a
\(d\)-dimensional space. The index function
\(f(XA): \mathbb{R}^{n \times d} \to \mathbb{R}\) defines a statistic to
measure a pattern of interest in the projected data, such as deviation
from normality, presence of clusters, or non-linear structure. For a
fixed sample of data, PP finds the orthonormal matrix \(A\) (also called
\emph{projection basis} in PP literature) that maximizes the index value
of the projection:

\begin{equation}\phantomsection\label{eq-optimization}{
\underset{A \in \mathcal{A}}{\max } \quad f(XA) \quad \text{subject to} \quad A'A = I_d
}\end{equation}

where \(I_d\) is the identity matrix of dimension \(d\).\\
It is interesting to note that when using PP visually, one cares less
about \(A\) than the plane described by \(A\), because the orientation
in the plane is irrelevant. The space of planes belongs to a Grassmann
manifold. This is usually how the projection pursuit guided tour (PPGT)
(Cook et al. 1995) operates, when using geodesic interpolation between
starting and target planes. It interpolates plane to plane, removing
irrelevant within-plane spin, and is agnostic to the basis (\(A\)) used
to define the plane. Thus, indexes which are used for the PPGT should be
rotationally invariant.

\begin{figure}

\centering{

\includegraphics[width=0.8\linewidth,height=\textheight,keepaspectratio]{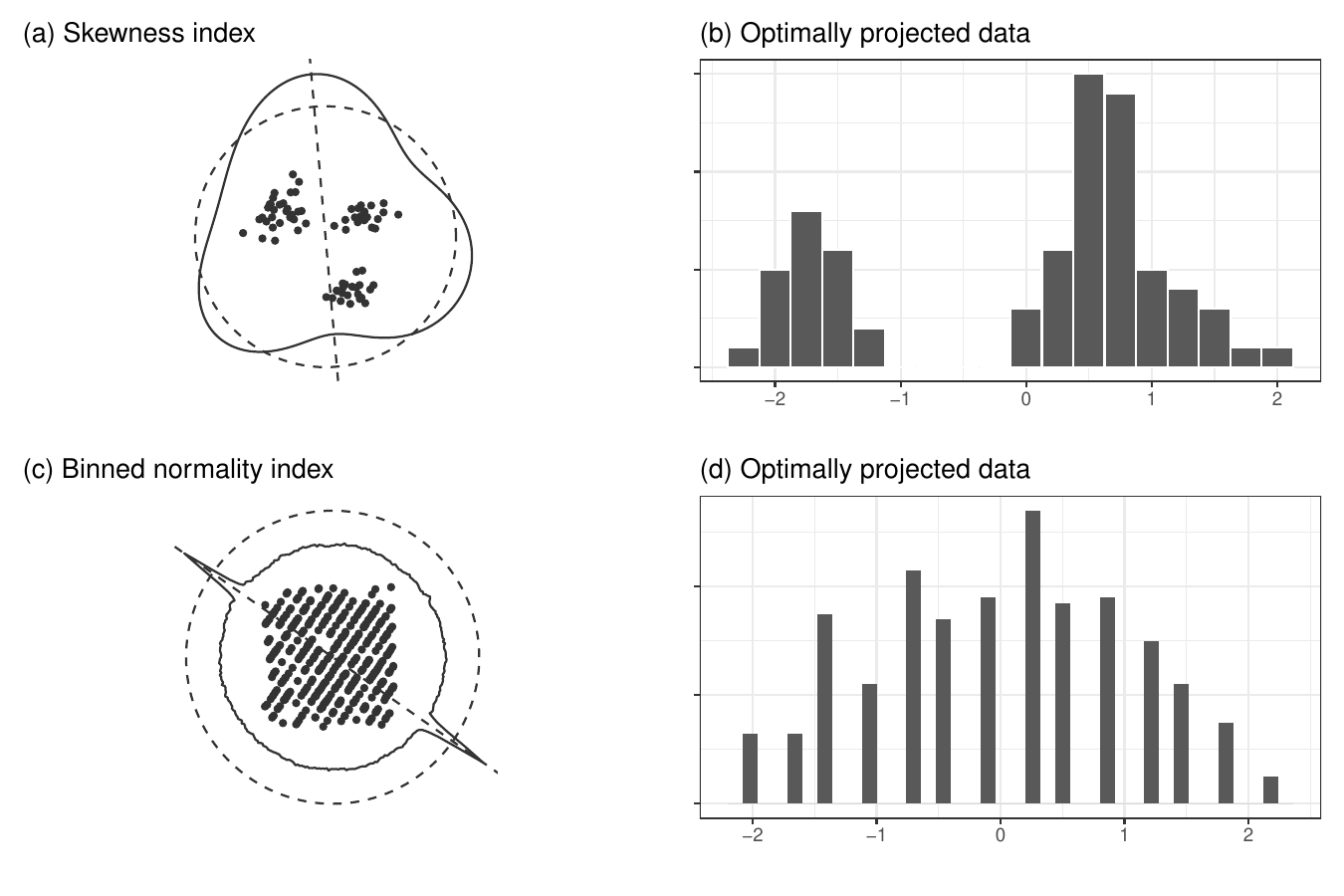}

}

\caption{\label{fig-example-functions}Examples of PP indexes with large
(top row) and small (bottom row) squint angles, shown with a Huber plot,
and histogram of the projected data corresponding to the optimal
projection. A Huber plot shows the PP index values for all 1D data
projections in polar coordinates. The skewness index on the trimodal
data is also smoother than the binned normality index on RANDU.}

\end{figure}

Index functions are quite varied in form, partially depending on the
data that is being projected. Figure~\ref{fig-example-functions} shows
two examples. Huber plots (Huber 1990) of 2D data sets are in (a) and
(c), showing the PP index values for all 1D projections of the 2D data
in polar coordinates, which reveals the form of these functions. The
dashed circle is a baseline set at the average value, and the straight
line marks the optimal projection. Plots (b) and (d) show the respective
best projections of the data as histograms. Indexes like the holes,
central mass and skewness (Cook, Buja, and Cabrera 1993) are generally
smooth for most data sets, but capture only large patterns. Many indexes
are noisy and non-convex, requiring an effective and efficient
optimization procedure to explore the data landscape and achieve a
globally optimal viewpoint of the data. The skewness index computed for
trimodal data, in (a), is smooth with a large squint angle but has three
modes, and thus is not convex. The binned normality index (a simple
version of a non-normality index as described in Huber 1985) computed on
the famous RANDU data, in (c), is noisier and has a very small squint
angle. The discreteness cannot be seen unless the optimizer is very
close to the optimal projection.

Optimization of PP is sometimes discussed when new indexes are proposed
(Posse 1995; Marie-Sainte, Berro, and Ruiz-Gazen 2010; Grochowski and
Duch 2011). Cook et al. (1995) tied the optimization closely to the
index with the introduction of the PPGT, which monitors the optimization
visually so that the user can see the projected data leading in and out
of the optimum. An implementation is available in the \texttt{tourr}
package (Wickham et al. 2011) in R (R Core Team 2023). Zhang et al.
(2021) illustrated how to diagnose optimization processes, particularly
focusing on the guided tour, and revealed a need for improved
optimization. While improving the quality of the optimization solutions
in the tour is essential, it is also important to be able to view the
data projections as the optimization progresses. Integrating the guided
tour with a global optimization algorithm that is efficient in finding
the global optimal and enables viewing of the projected data during the
exploration process is a goal.

In this work, the potential for a Jellyfish Search Optimizer (JSO) (Chou
and Truong 2021; Rajwar, Deep, and Das 2023) for the PPGT is explored.
The JSO, inspired by the search behavior of jellyfish in the ocean, is a
swarm-based metaheuristic designed to solve global optimization
problems. Compared to other metaheuristic methods, JSO has demonstrated
stronger search ability and faster convergence, and requires fewer
tuning parameters. These practicalities make JSO a promising candidate
for enhancing PP optimization.

The primary goal of the study reported here is to investigate the
performance of JSO in PP optimization for the guided tour. It is of
interest to assess how quickly and closely the optimizer reaches a
global optimum, for various PP indexes that may have differing
complexities. To observe the performance of JSO with different types of
PP indexes, metrics are introduced to capture specific properties of the
index including squintability (based on Tukey and Tukey 1981's squint
angle) and smoothness. We mathematically define metrics for
squintability and smoothness, which is a new contribution for PP
research. A series of simulation experiments are conducted using JSO to
detect different target patterns (pipe and sine-wave) with different PP
indexes (holes, MIC, TIC, dcor, loess, splines, skinny, and stringy), as
well as JSO's hyper-parameter choices (number of jellyfish and maximum
number of iterations). This work should facilitate better optimization
for PP and guide the choice of optimizer when designing new PP indexes.

The paper is structured as follows. Section~\ref{sec-background}
introduces the background of the PPGT, reviews existing optimizers and
index functions in the literature. Section~\ref{sec-PP-properties}
introduces the metrics that measure two properties of PP indexes,
smoothness and squintability. Section~\ref{sec-JSO} describes the new
JSO for PP. Section~\ref{sec-sim-deets} outlines two simulation
experiments to assess JSO's performance: one comparing JSO's performance
improvements relative to an existing optimizer, Creeping Random Search
(CRS), and the other studying the impact of PP index properties,
controlling for data dimension and JSO hyper-parameters, on optimization
performance, and Section~\ref{sec-sim-res} presents the results.
Section~\ref{sec-discussion} discusses the implementation of JSO in the
\texttt{tourr} package and the PP property calculation in the
\texttt{ferrn} package. Section~\ref{sec-conclusion} summarizes the work
and provides suggestions for future directions.

\section{Projection pursuit, tours, index functions and
optimization}\label{sec-background}

A tour on high-dimensional data is constructed by geodesically
interpolating between pairs of planes. Any plane is described by an
orthonormal basis, \(A_t\), where \(t\) represents time in the sequence.
The term ``geodesic'' refers to maintaining the orthonormality
constraint so that each view shown is correctly a projection of the
data. The PP guided tour operates by geodesically interpolating to
target planes (projections) which have high PP index values, as provided
by the optimizer. The geodesic interpolation means that the viewer sees
a continuous sequence of projections of the data, so they can watch
patterns of interest forming as the function is optimized. There are
five optimization methods implemented in the \texttt{tourr} package:

\begin{itemize}
\tightlist
\item
  a pseudo-derivative, that searches locally for the best direction,
  based on differencing the index values for very close projections.
\item
  a brute-force optimization (CRS).
\item
  a modified brute force algorithm described in Posse (1995).
\item
  an essentially simulated annealing (Bertsimas and Tsitsiklis 1993)
  where the search space is reduced during the optimization.
\item
  a very localized search, to take tiny steps to get closer to the local
  maximum.
\end{itemize}

There are numerous PP index functions available: introduced in Huber
(1985), Cook, Buja, and Cabrera (1993), Lee et al. (2005), Lee and Cook
(2010), Grimm (2016), Laa et al. (2022). Most are relatively simply
defined, for any projection dimension, and implemented because they are
relatively easy to optimize. A goal is to develop PP indexes based on
scagnostics (Wilkinson, Anand, and Grossman 2005; Wilkinson and Wills
2008), but the blockage is their optimization as these tend to be noisy,
with potentially small squint angles.

An initial investigation of PP indexes, and the potential for
scagnostics is described in Laa and Cook (2020). To be useful here an
optimizer needs to be able to handle index functions that are possibly
not very smooth. In addition, because the target structure in the data
might be relatively fine, the optimizer needs to be able to find maxima
that occur with a small squint angle, that can only be seen from very
close by. One last aspect that is useful is for an optimizer to return
local maxima in addition to the global one because data can contain many
different and interesting features.

\section{Properties of PP indexes}\label{sec-PP-properties}

Laa and Cook (2020) has proposed five criteria for assessing projection
pursuit indexes (smoothness, squintability, flexibility, rotation
invariance, and speed). Since not all index properties affect the
optimization process, the focus of this work is on the first two
properties, \emph{smoothness} (Section~\ref{sec-smoothness}) and
\emph{squintability} (Section~\ref{sec-squintability}), for which
metrics are proposed to quantify them.

\subsection{Smoothness}\label{sec-smoothness}

A classical way to describe the smoothness of a function is to identify
how many continuous derivatives of the function exist. This can be
characterized by Sobolev spaces (Adams and Fournier 2003).

\begin{defn}[Sobolev space]\label{def:sobolev_space}
The Sobolev space $W^{k,p}(\mathbb{R})$ for $1\leq p\leq \infty$ is the set of all functions $f$ in $L^p(\mathbb{R})$ for which all weak derivatives $f^{(\ell)}$ of order $\ell\leq k$ exist and have a finite $L^p$ norm.
\end{defn}

The Sobolev index \(k\) in Definition \ref{def:sobolev_space} can be
used to characterize the smoothness of a function: if \(f\in W^{k,p}\),
then the higher \(k\), the smoother \(f\). While this Sobolev index
\(k\) is a useful measure of smoothness, it can be difficult to compute
or even estimate in practice. For example, the loess and splines indexes
are not differentiable due to the use of the \(\max\) operator in their
definitions. Scagnostic indexes, such as skinny and stringy, are defined
based on graph elements (area, diameter, perimeter, and length) of alpha
hull or minimum spanning tree).

To obtain a computable estimator of the smoothness of the index function
\(f\), we propose an approach based on random fields. If a PP index
function \(f\) is evaluated at some random bases, then these random
index values can be interpreted as a random field, indexed by a space
parameter, namely the random projection basis. This analogy suggests to
use this random training sample to fit a spatial model. We propose to
use a Gaussian process equipped with a Matérn covariance function, due
to the connections between this model and Sobolev spaces, see for
example Porcu et al. (2024).

The distribution of a Gaussian process is fully determined by its mean
and covariance function. The smoothness property comes into play in the
definition of the covariance function: if a PP index is very smooth,
then two close projection bases should produce close index values
(strong correlation); by contrast, if a PP index is not very smooth,
then two close projection bases might give very different index values
(fast decay of correlations with respect to distance between bases).
Popular covariance functions are parametric positive semi-definite
functions. In particular, the Matérn class of covariance functions has a
dedicated parameter to capture the smoothness of the Gaussian field.

\begin{defn}[Matérn covariance function]\label{def:matern}
The Matérn covariance function $K$ is defined by
\begin{equation}
K(u)=K_{\nu,\eta,\ell}(u):=\eta^2\frac{\left(\sqrt{2\nu}\frac{\left\Vert u\right\Vert}{\ell}\right)^{\nu}}{\Gamma(\nu)2^{\nu-1}}\mathcal{K}_{\nu}\left(\sqrt{2\nu}\frac{\left\Vert u\right\Vert}{\ell}\right)\ ,\label{eq:matern}
\end{equation}
where $\left\Vert u\right\Vert$ is the Euclidean norm of $u\in\mathbb{R}^{p{\times}d}$, $\nu>0$ is the smoothness parameter, $\eta$ is the outputscale, $\ell$ is the lengthscale, and $\mathcal{K}_\nu$ is
the modified Bessel function [DLMF 10.25].
\end{defn}

\begin{figure}

\centering{

\includegraphics[width=1\linewidth,height=\textheight,keepaspectratio]{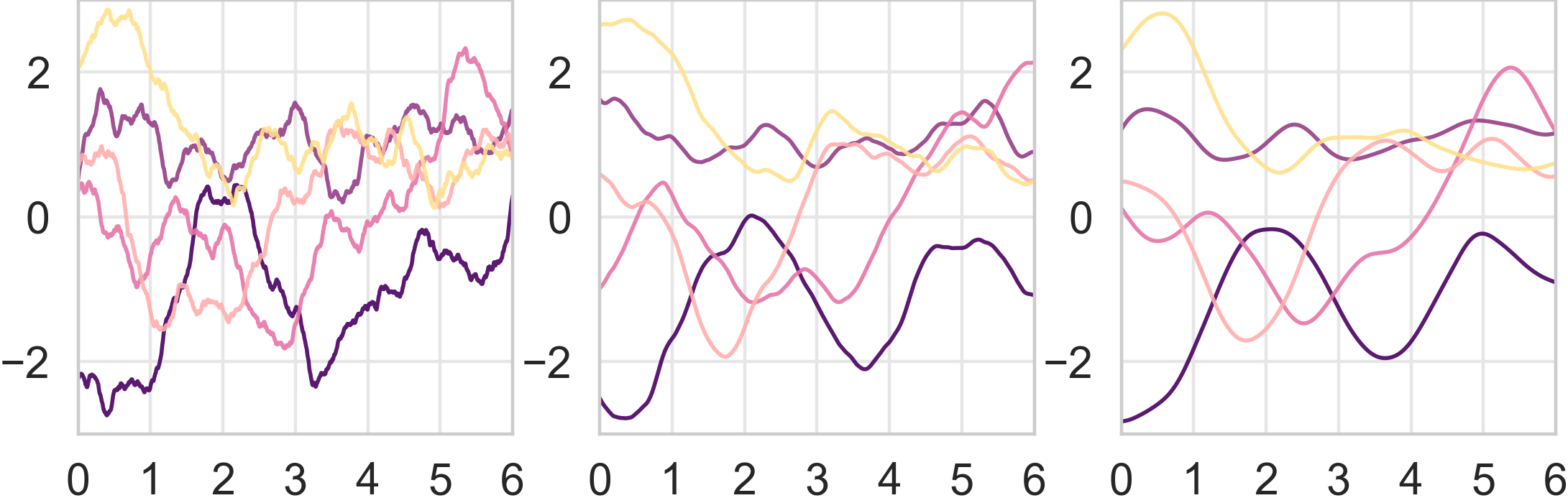}

}

\caption{\label{fig-matern-1d}Five random simulations from a Gaussian
Process defined on \(\mathbb{R}\) with zero mean and Matérn-\(\nu\)
covariance function, with \(\nu=1\) (left), \(\nu=2\) (middle), and
\(\nu=4\) (right), showing that higher values of \(\nu\) produce
smoother curves.}

\end{figure}

\begin{figure}

\centering{

\includegraphics[width=1\linewidth,height=\textheight,keepaspectratio]{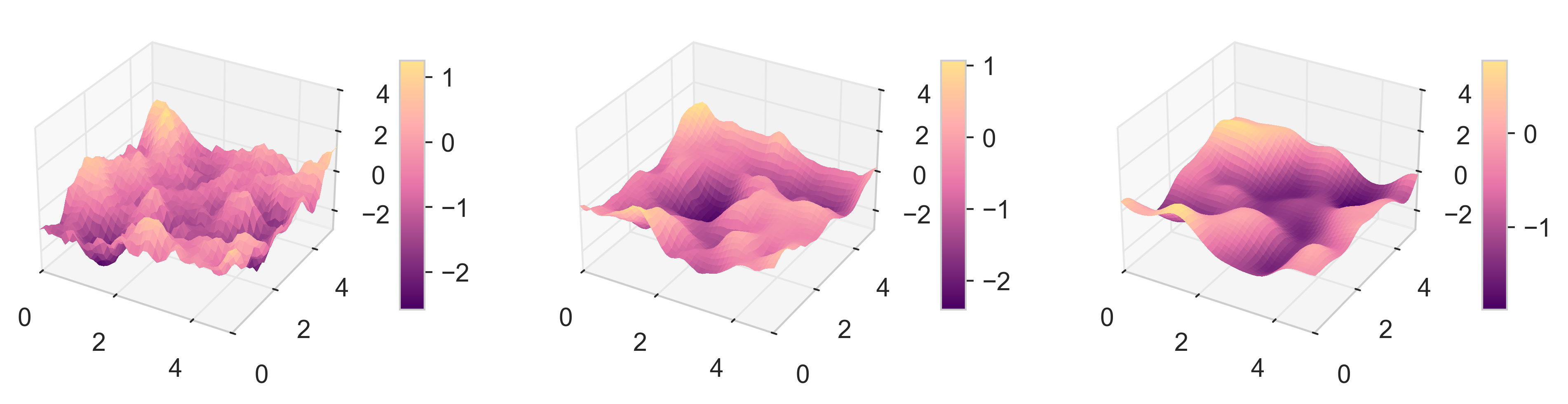}

}

\caption{\label{fig-matern-2d}One random simulation from a Gaussian
Process defined on \(\mathbb{R}^2\) with zero mean and Matérn-\(\nu\)
covariance function, with \(\nu=1\) (left), \(\nu=2\) (middle), and
\(\nu=4\) (right), showing that higher values of \(\nu\) produce
smoother surfaces.}

\end{figure}

The Matérn covariance function can be expressed analytically when
\(\nu\) is a half-integer, the most popular values in the literature
being \(\frac{1}{2}\), \(\frac{3}{2}\) and \(\frac{5}{2}\) (Rasmussen
and Williams 2006). The parameter \(\nu\), called \emph{smoothness
parameter}, controls the decay of the covariance function. As such, it
is an appropriate measure of smoothness of a random field, as shown by
the simulations on Figure~\ref{fig-matern-1d} and
Figure~\ref{fig-matern-2d}. For example, Karvonen (2023) showed that if
a function \(f\) has a Sobolev index of \(k\), then the smoothness
parameter estimate \(\nu\) in \eqref{eq:matern} cannot be asymptotically
less than \(k\). See the survey Porcu et al. (2024) for additional
results on the connection between the Matérn model and Sobolev spaces.
An interesting result is that the asymptotic case
\(\nu\rightarrow\infty\) coincides with the Gaussian kernel:
\(K_\infty(u)=\exp(-{\left\Vert u\right\Vert}^{2}/2)\).

In view of these results, the parameter \(\nu\) is suggested as a
measure of the smoothness of the PP index function by fitting a Gaussian
process prior with Matérn covariance on a dataset generated by random
evaluations of the index function. There exist several R packages, such
as \texttt{GpGp} (Guinness, Katzfuss, and Fahmy 2021) or
\texttt{ExaGeoStatR} (Abdulah et al. 2023), to fit the hyper-parameters
of a GP covariance function on data, which is usually done by maximum
likelihood estimation. In this project, the \texttt{GpGp} package is
used.

\begin{defn}
Let $\mathbf{A}=[A_1, \ldots, A_N] \in (\mathbb{R}^{p \times d})^N$ be d-dimensional projection bases, $\mathbf{y}=[f(XA_1),\ldots,f(XA_N)]$ be the corresponding PP index values, and $\mathbf{K}=[K_\theta(A_{i}-A_{j})]_{1\leq i,j\leq N}\in\mathbb{R}^{N\times N}$ be the Matérn covariance matrix evaluated at the input bases, where the vector $\theta$ contains all the parameters of the multivariate Matérn covariance function $K$ (smoothness, outputscale, lengthscales). The log-likelihood of the parameters $\theta$ is defined by 
\begin{equation}
\mathcal{L}(\theta)=\log p(\mathbf{y}\left|\mathbf{A},\theta\right.)=-\frac{1}{2}\mathbf{y}^{\top}(\mathbf{K}+\sigma^{2}\mathbf{I})^{-1}\mathbf{y}-\frac{1}{2}\mathrm{\log}(\det(\mathbf{K}+\sigma^{2}\mathbf{I}))-\frac{N}{2}\log(2\pi)\, \label{eq:gp_log_likelihood}
\end{equation}
where the nugget parameter $\sigma$ is the standard deviation of the intrinsic noise of the Gaussian process.
The optimal parameters are obtained by maximum log-likelihood
\begin{equation}
\theta^* = \underset{\theta}{\max}\mathcal{L}(\theta)
\end{equation}
The resulting optimal smoothness parameter $\nu$ is chosen as our smoothness metric.
\end{defn}

The value of the optimal smoothness parameter \(\nu>0\) can be naturally
interpreted as follows: the higher \(\nu\), the smoother the index
function.

\subsection{Squintability}\label{sec-squintability}

The squintability metric is inspired by the concept of squint angle in
Definition \ref{def:squint-angle}, originally introduced in Tukey and
Tukey (1981) and later discussed in Laa and Cook (2020). Consider the
space of all projection bases: those near the optimal basis correspond
to projections with structure of interest. The size of this neighborhood
depends on the choice of the index function. For some indexes,
projections resemble the target even when relatively far from the
optimum, corresponding to a large squint angle, while for others,
projections need to be close to the optimum to reveal structure we are
interested, resulting in a small squint angle. For PP, a small squint
angle is considered to be undesirable because it means that the
optimizer needs to sample into a very narrow region to be able to
``see'' the optimum, making it difficult for the optimizer to find the
optimum.

\begin{defn}[squint angle]\label{def:squint-angle}
Let $A$ and $B$ be two $d$-dimensional orthonormal matrices in $\mathbb{R}^p$. The squint angle $\theta$ between the subspace spanned by $A$ and $B$ is defined as the smallest principal angle between these subspaces: $\theta = \min_{i \in \{1, \cdots, d\}} \arccos(\tau_i)$, where $\tau_i$ are the singular values of the matrix $M = A^T B$ obtained from its singular value decomposition.
\end{defn}

To quantify how close projections need to be before structures of
interest can be observed (the optimizer starts hill-climbing), we can
track how the index value changes as we move from an initial projection
basis towards the optimal one, that is, how index value changes as the
projection distance becomes smaller.

\begin{defn}[projection distance]\label{def:proj-dist}
Let $A \in \mathbb{R}^{p \times d}$ be a $d$-dimensional orthonormal matrix, and let $A^*$ be the optimal matrix that achieves the maximum index value for a given data. The projection distance between $A$ 
and $A^*$, $r(A, A^*)$, is defined by
$r(A, A^*) = \lVert AA^\prime - A^*A^{*\prime}\,\rVert _F$
where $\lVert . \rVert _F$ denotes the Frobenius norm, given by
$\lVert M \rVert _F = \sqrt{\sum_{ij} M_{ij}^2}$. 
\end{defn}

\begin{figure}

\centering{

\pandocbounded{\includegraphics[keepaspectratio]{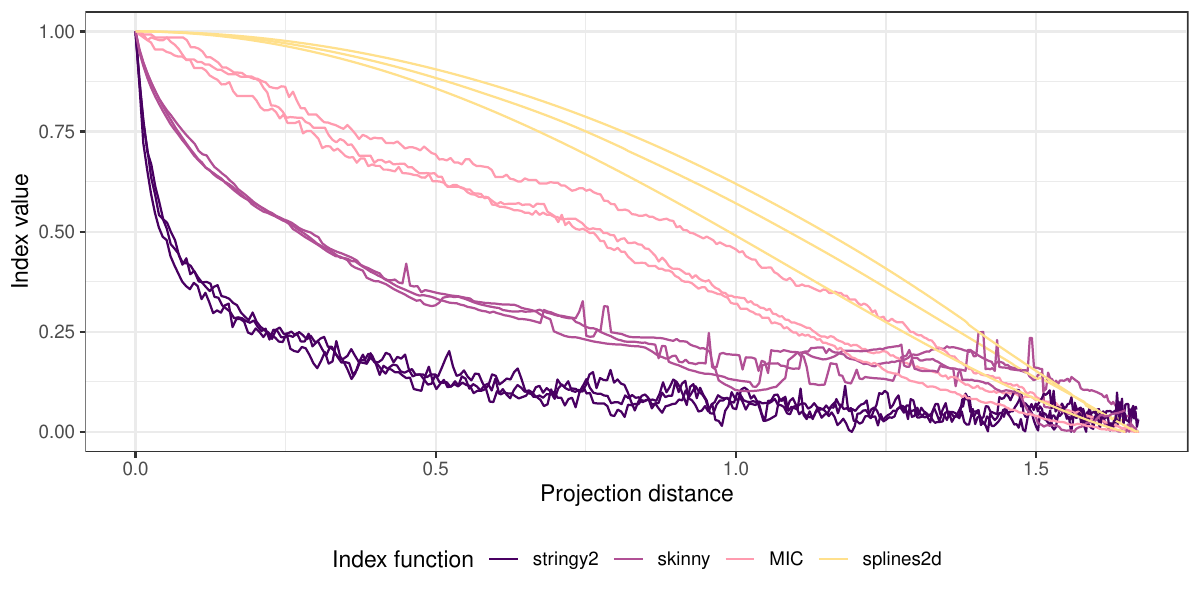}}

}

\caption{\label{fig-squintability-ill}Simulated traces of index value as
a function of projection distances for four PP index functions: MIC,
splines2d, skinny, and stringy2. The index function MIC and splines2d
make early progression towards the optimal index value, indicating a
large squint angle, whereas the traces from skinny, and specially
stringy2, show improvement only near the optimal projection, suggesting
low squintability.}

\end{figure}%

Figure~\ref{fig-squintability-ill} illustrates some simulated traces for
different PP index functions: MIC, splines2d, skinny, and stringy2. It
is expected that for a PP index with high squint angle, the optimization
\eqref{eq-optimization} should make substantial progress early on.
Conversely, for a PP index with low squint angle, it might take a long
while for the optimization to make substantial progress, as the
candidate projections would need to be very close to the optimal one for
the structure of the index function to be visible enough to be amenable
to efficient optimization. These empirical observations suggest that the
half-point index value improvement percentage provides an appropriate
mathematical definition of squintability, which matches the intuition
behind this concept, while being amenable to numerical computation.

\begin{defn}[squintability]\label{def:squintability}
Let $g: \mathbb{R} \mapsto  \mathbb{R}$ be a decreasing function that models the index value $f(XA)$ as a function of the projection distance $r(A, A^*)$, such that $g(r) = g(r(A, A^*)) \simeq f(XA)$.  The squintability of an index function $f$ is then defined by 

\begin{equation}
\varsigma(f) = \frac{g(r_{0}/2)-g(r_{0})}{g(0)-g(r_{0})} \in [0,1]
\label{eq-squintability}
\end{equation}

\end{defn}

where \(r_0 = r(A_0, A^*)\) is the projection distance from the initial
projection \(A_0\) to the optimal one \(A^*\). Remark that the amount by
which the index value can improve from the starting matrix \(A_0\) to
the optimal one \(A^*\) is given by \(g(0)-g(r_{0})\). As a result,
equation~\eqref{eq-squintability} represents the proportion of this
maximum improvement which has been achieved by the time the distance
\(r_0\) to the optimal matrix has been reduced by half (\(r_0/2\)).

To compute the squintability metric \eqref{eq-squintability} in
practice, several approaches are possible. The first one is to propose a
parametric model for \(g\), and use it to obtain an explicit formula for
\(\varsigma\). Numerical experiments suggest a scaled sigmoid shape as
described below. Define
\begin{equation}\phantomsection\label{eq-logistic}{
\ell(x):=\frac{1}{1+\exp(\theta_{3}(x-\theta_{2}))}\ ,
}\end{equation}

which is a decreasing logistic function depending on two parameters
\(\theta_2\) and \(\theta_3\), such that
\(\ell(\theta_{2})=\frac{1}{2}\). Then, define
\begin{equation}\phantomsection\label{eq-parametric}{
g(x)=(\theta_{1}-\theta_{4})\frac{\ell(x)-\ell(r_0)}{\ell(0)-\ell(r_0)}+\theta_{4}\ ,
}\end{equation}

which depends on three additional parameters, \(\theta_1\),
\(\theta_4\), and \(r_0\), such that \(g(0)=\theta_1\) and
\(g(r_0)=\theta_4\). Under the parametric model (\ref{eq-parametric}),
the squintability metric \eqref{eq-squintability} can be shown to be
equal to
\begin{equation}\phantomsection\label{eq-squintability-parametric}{
\varsigma=\frac{g\left(r_{0}/2\right)-\theta_{4}}{\theta_{1}-\theta_{4}}=\frac{\ell(r_0/2)-\ell(r_0)}{\ell(0)-\ell(r_0)}\ .
}\end{equation}

In practice, the parameters of this model (\ref{eq-parametric}) can be
estimated numerically, for example by non-linear least squares, and then
used to evaluate \(\varsigma\) as in equation
(\ref{eq-squintability-parametric}).

Alternatively, one can estimate \eqref{eq-squintability} in a
nonparametric way, for example by fitting \(g\) using kernel regression,
then numerically estimate \(\varsigma\) from its definition
\eqref{eq-squintability}.

\section{The jellyfish search optimizer}\label{sec-JSO}

The Jellyfish Search Optimizer (JSO) mimics the natural movements of
jellyfish, which include passive and active motions driven by ocean
currents and their swimming patterns, respectively. In the context of
optimization, these movements are abstracted to explore the search
space, aiming to balance exploration (searching new areas) and
exploitation (focusing on promising areas). The algorithm aims to find
the optimal solution by adapting the behavior of jellyfish to navigate
towards the best solution over iterations (Chou and Truong 2021).

To solve the optimization problem embedded in the PPGT with JSO, a
starting projection, an index function, the number of jellyfish, and the
maximum number of iterations are provided as input. Then, the current
projection is evaluated by the index function. The projection is then
moved in a direction determined by a random time control function,
\(c_t\), whose value decreases as the number of trials increases
(subject to randomness) to guide the exploration and exploitation phases
of the algorithm. When \(c_t \ge 0.5\), new directions will be taken
like a jellyfish explores with the ocean trend, defined as the
difference of the current best jellyfish and average of all the current
jellyfish. When \(1- c_t\) is smaller than a randomly generated number
in \([0, 1]\), jellyfish will move passively with small random
perturbations, otherwise, it will move actively towards or away from
another jellyfish based on their index values. A new projection is
accepted if it is an improvement compared to the current one, rejected
otherwise. This process continues and iteratively improves the
projection, until the pre-specified maximum number of trials is reached.

\begin{tcolorbox}[enhanced jigsaw, bottomtitle=1mm, breakable, colbacktitle=quarto-callout-note-color!10!white, opacitybacktitle=0.6, title={Algorithm: Jellyfish Optimizer Pseudo Code}, left=2mm, arc=.35mm, rightrule=.15mm, colframe=quarto-callout-note-color-frame, colback=white, leftrule=.75mm, bottomrule=.15mm, toptitle=1mm, toprule=.15mm, titlerule=0mm, opacityback=0, coltitle=black]

\textbf{Input}: \texttt{current\_projections}, \texttt{index\_function},
\texttt{trial\_id}, \texttt{max\_trial}

\textbf{Output}: \texttt{optimized\_projection}

\textbf{Initialize} \texttt{current\_best} as the projection with the
best index value from \texttt{current\_projections}, and
\texttt{current\_idx} as the array of index values for each projection
in \texttt{current\_projections}

\textbf{for} each \texttt{trial\_id} in 1 to \texttt{max\_tries}
\textbf{do}

\begin{quote}
Calculate the time control value, \(c_t\), based on \texttt{trial\_id}
and \texttt{max\_trial}
\end{quote}

\begin{quote}
\textbf{if} \(c_t\) is greater than or equal to \(0.5\) \textbf{then}

\begin{quote}
Define the ocean current trend as the difference of the
\texttt{current\_best} and the average of the
\texttt{current\_projections}
\end{quote}

\begin{quote}
Update each projection towards the ocean trend using a random factor and
orthonormalisation
\end{quote}

\textbf{else}

\begin{quote}
\textbf{if} a random number is greater than \(1 - c_t\) \textbf{then}

\begin{quote}
Slightly adjust each projection with a small random factor (passive)
\end{quote}

\textbf{else}

\begin{quote}
For each projection, compare with a random jellyfish in
\texttt{current\_projections} and adjust towards or away from it based
on their corresponding \texttt{current\_idx} (active)
\end{quote}
\end{quote}

Update the orientation of each projection to maintain consistency

Evaluate the new projections using the index function
\end{quote}

\begin{quote}
\textbf{if} any new projection is worse than the current, revert to the
\texttt{current\_projections} for that case

\begin{quote}
Determine the projection with the best index value as the new
\texttt{current\_best}
\end{quote}
\end{quote}

\begin{quote}
\textbf{exit}
\end{quote}

\textbf{return} the set of projections with the updated
\texttt{current\_best} as the \texttt{optimized\_projection}

\end{tcolorbox}

The JSO implementation involves several key parameters that control its
search process in optimization problems. While the specific
implementation details can vary depending on the version of the
algorithm or its application, the focus is on two main parameters that
are most relevant to our application: the number of jellyfish and the
maximum number of iterations.

\section{Assessing the optimizers}\label{sec-sim-deets}

This section explains the details of two simulation studies: (1) a
comparison between the JSO and an existing optimizer, creeping random
search (Zhang et al. 2021; Laa and Cook 2020), and (2) a modelling study
to examine how factors (smoothness, squintability, data dimension, and
JSO hyper-parameters) affect its success rate in detecting holes and
sine-wave pattern across a collection of index functions.

\subsection{Performance of JSO relative to CRS}\label{sec-app-1}

The CRS is the main optimization routine currently used for the guided
tour. Here the two optimizers are compared on the task of finding the 2D
pipe structure using the \texttt{holes} index in data at dimensions 6,
8, 10, and 12. Additional variables are Gaussian noise. JSO uses 100
jellyfish with 100 iterations, while CRS allows up to 1000 trials per
iteration and terminates if no better projection is found within these
trials. These choices enable fair comparison between CRS and JSO, that
conforms to how they are used in practice. Each task is repeated 50
times to evaluate the performance between the two optimizers.

\begin{figure}[!ht]

\centering{

\includegraphics[width=5.41in,height=\textheight,keepaspectratio]{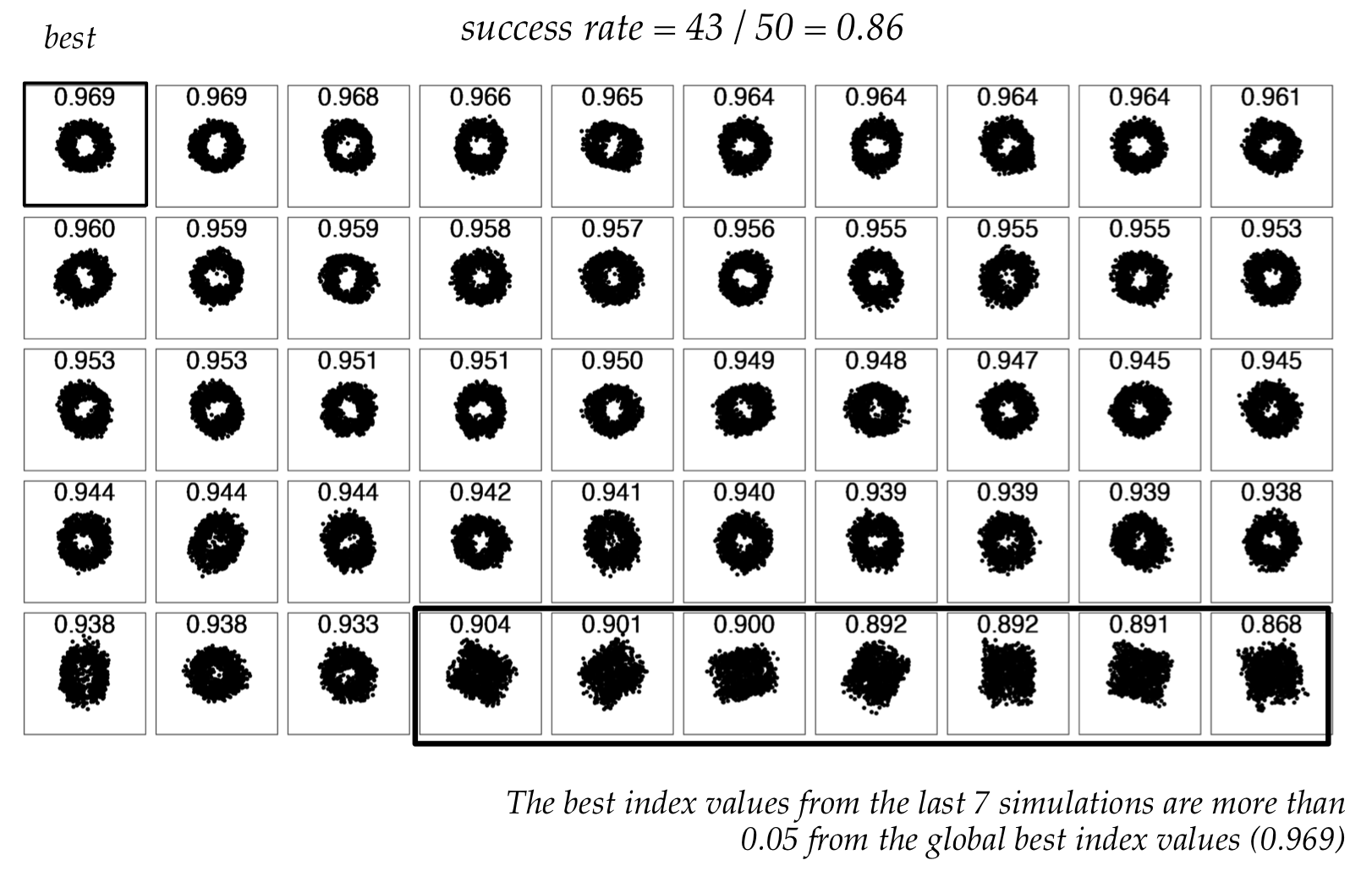}

}

\caption{\label{fig-success-rate}How success rate is calculated,
illustrated using the optimal projections from 50 optimizations of 8D
pipe data, optimized by CRS, sorted by index value. The pipe shape is
recognizable in the projection index values between 0.933-0.969. Of the
50 simulations, 43 achieved an index value within 0.05 of the best,
resulting in a success rate of 0.86.}

\end{figure}

The performance of the optimizers is measured by the success rate, which
is defined as the proportion of simulations that achieves a final index
value within 0.05 of the best index value found among all 50
simulations. Figure~\ref{fig-success-rate} illustrates why the choice of
0.05 is reasonable: the pipe is not recognizable in the projected data.
This is motivated by Laa and Cook (2020)'s approach to investigating PP
indexes.

The results of the simulation are collected using the data structure
used in Zhang et al. (2021) for assessing PP optimizers. The design
parameters are stored along with index value, projection basis, and
random seed used.

\subsection{Factors affecting JSO success rate: JSO hyper-parameters and
index properties}\label{sec-app-2}

To examine the performance of JSO relative to hyper-parameter choices,
the \texttt{holes} index is used to find the 2D pipe in data of
dimensions 4, 6, 8, 10, and 12. The levels of hyper-parameters are 20,
50, and 100 jellyfish and a maximum of 50 and 100 attempts. Each task is
repeated fifty times to calculate the success rate as described in
Section~\ref{sec-app-1}.

This set of simulations is expanded to assess the performance of JSO
relative to the index properties, with a second data set type and
additional PP indexes. The second data set has a sine wave in two
dimensions and Gaussian noise in the remaining dimensions. Seven
additional PP indexes that should be able to extract the sine wave are
used (dcor, loess, MIC, TIC, splines, skinny and stringy2) following Laa
and Cook (2020). (The supplementary material includes definitions of
these PP indexes, and descriptions of the pipe and sine wave data
simulation.) The JSO hyper-parameters, number of jellyfish and maximum
iterations, are also included. This results in a total of 77 scenarios,
comprising 30 computed on the pipe data and 47 on the sine wave data.
The choice of 4-12 dimensions covers the range of easy to relatively
difficult for the optimizers to detect the structured projections. The
same approach is used, that each experiment is repeated 50 times and
success rate is calculated at each level of the simulation design.

\begin{figure}

\centering{

\includegraphics[width=1\linewidth,height=\textheight,keepaspectratio]{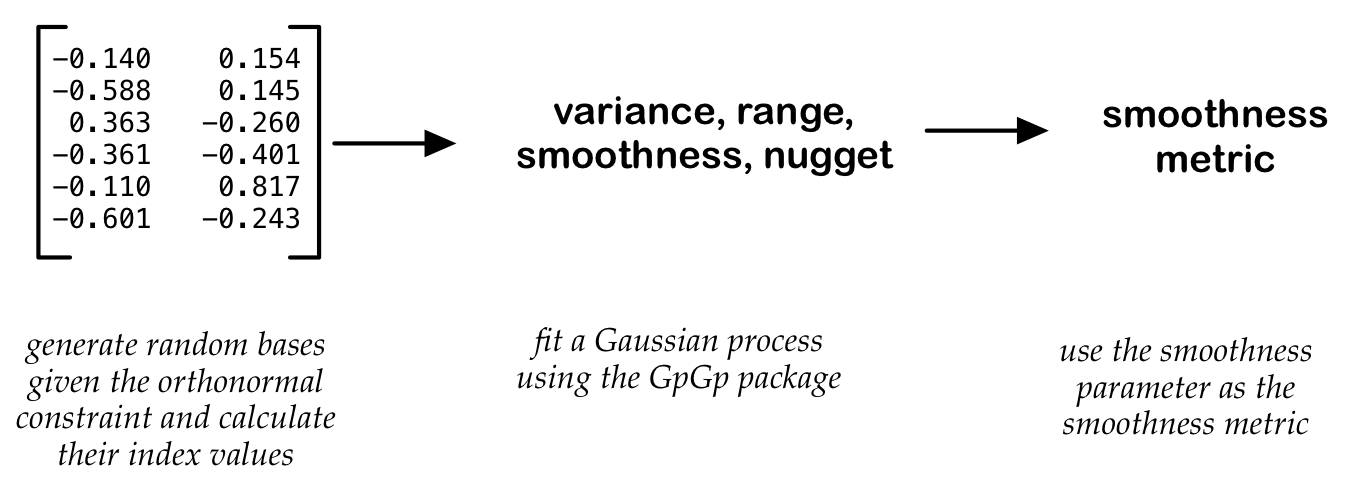}

}

\caption{\label{fig-smoothness}Steps for calculating smoothness in a
projection pursuit problem: 1) sample random bases given the
orthonormality contraint and calculate their corresponding index values,
2) fit a Gaussian process model of index values against the bases to
obtain the smoothness measure, 3) take the smoothness parameter from the
Gaussian process as the smoothness metric.}

\end{figure}%

Recall that smoothness and squintability are metrics used to
characterize the index function for PP tasks and are invariant to the
optimizer and its hyper-parameters. Thus simulations with different JSO
hyper-parameters for the same PP task share the same smoothness value.
In total, smoothness and squintability metrics are calculated on the 23
PP tasks in our simulation.

Figure~\ref{fig-smoothness} describes the procedure to compute
smoothness. For each PP task, 500 random bases are simulated and index
values are calculated for each. A Gaussian process model is then fitted
to the resulting data to obtain the smoothness measure for the index, as
described in Section~\ref{sec-smoothness}.

\begin{figure}

\centering{

\includegraphics[width=1\linewidth,height=\textheight,keepaspectratio]{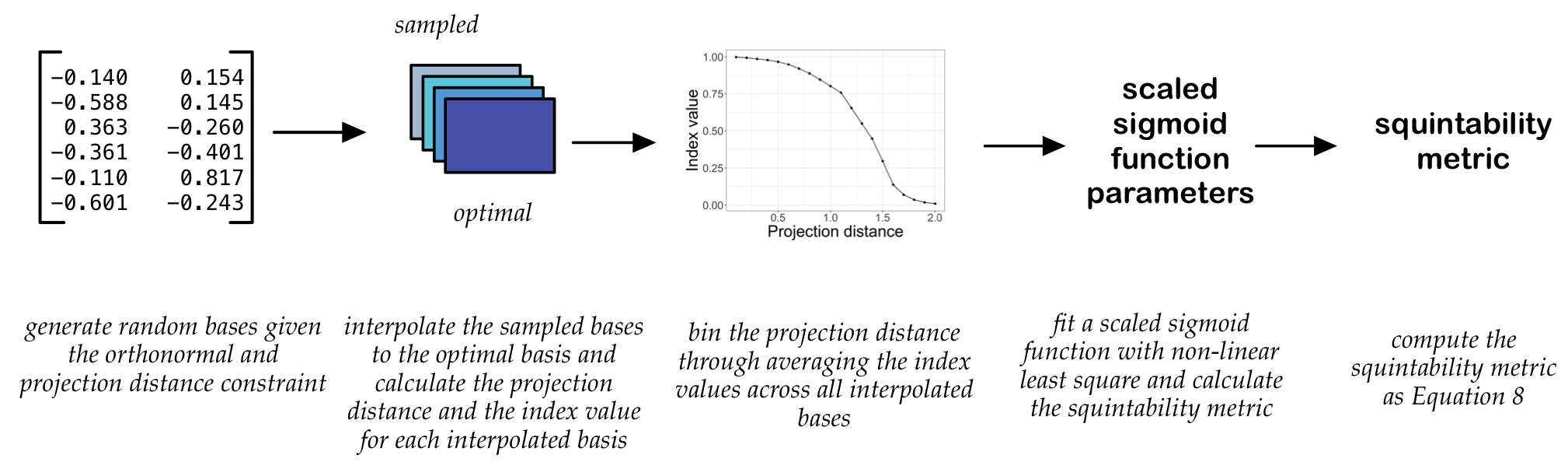}

}

\caption{\label{fig-squintability}Steps for calculating squintability in
a projection pursuit problem: 1) sample random bases given the
orthonormality and projection distance contraint and calculates their
corresponding index values, 2) interpolate the sampled bases to the
optimal basis and calculate the projection distance and the index value.
3) average the index values by projection distances at each 0.005
distance bin, 4) fit the scaled sigmoid function in Equation
\eqref{eq-logistic} and \eqref{eq-parametric} , 5) calculate the
squintability metric using Equation
(\ref{eq-squintability-parametric}).}

\end{figure}%

Figure~\ref{fig-squintability} describes the procedure to compute
squintability. For each PP task, 50 random bases are generated and
interpolated to the optimal basis with a step size of 0.005. The index
values are averaged over the distance window of 0.005. Index values are
calculated for all the interpolated bases and averaged over the distance
bin of 0.005. The logistic function in Equation (\ref{eq-logistic}) and
(\ref{eq-parametric}) is fitted with non-linear least squares and
calculate the squintability measure using Equation
(\ref{eq-squintability-parametric}).

A generalized linear model is fitted using a quasibinomial family and a
logit link function to assess the factors affecting the success rate.
Predictors are smoothness, squintability, data dimension, and JSO
hyper-parameters. For numerical stability related to small squintability
values associated with the stringy index (squintability less than
0.002), we decide to use the rank of the two metrics, rather than the
nominal values, in the model.

\section{Results}\label{sec-sim-res}

This section summarizes the findings from the simulations that compare
the JSO performance with the existing CRS optimizer, and the
relationship between optimization success and hyper-parameter choices
and PP index properties.

\begin{figure}[!ht]

\centering{

\pandocbounded{\includegraphics[keepaspectratio]{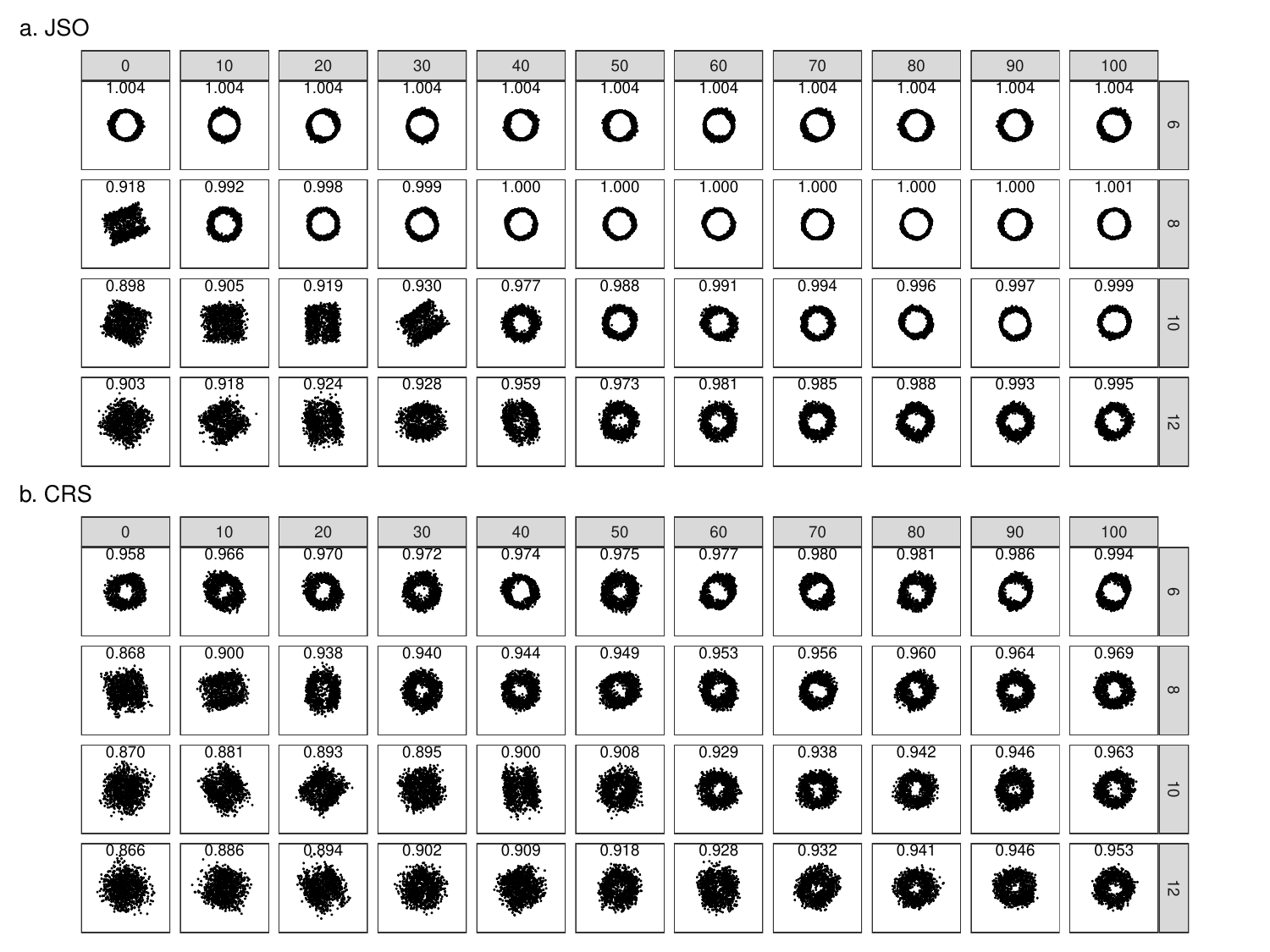}}

}

\caption{\label{fig-proj}Visual comparison of JSO and CRS results, using
optimal data projections obtained over 50 simulations of the pipe data.
Rows correspond to data dimension. Columns correspond to quantiles of
the index values, with 0 being the minimum, 100 being the maximum, and
50 the median. JSO achieves the better views of the pipe generally than
CRS. As dimension increases both have more difficulty finding the pipe.}

\end{figure}%

\subsection{Performance of JSO relative to
CRS}\label{performance-of-jso-relative-to-crs}

The optimal projection is compared between JSO and CRS for the pipe data
at dimensions 6, 8, 10, and 12 in Figure~\ref{fig-proj}. The columns
show the quantile of index value in the 50 optimal projections, with 0
being the minimum, 100 being the maximum, and 50 the median. The purpose
is to summarize the views of the data resulting from different
optimization, and hence compare results between the two optimizers.
Generally, the JSO does a more consistent job of finding the pipe
structure clearly. As the data dimension increases, both optimizers
struggle and less clearly capture the circle shape.

\subsection{Effect of hyper-parameters effect on JSO success
rate}\label{effect-of-hyper-parameters-effect-on-jso-success-rate}

\begin{figure}

\centering{

\pandocbounded{\includegraphics[keepaspectratio]{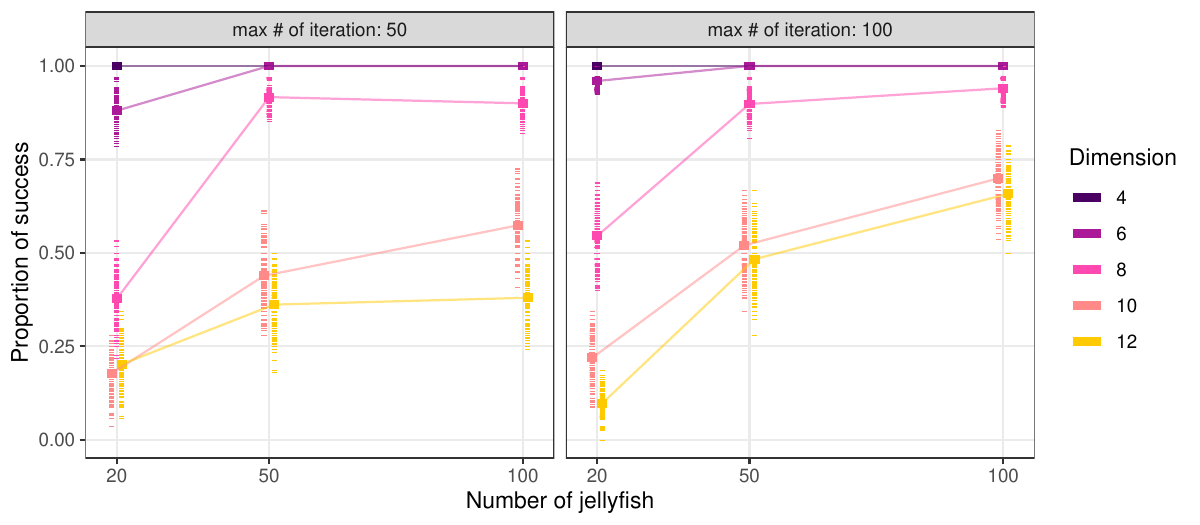}}

}

\caption{\label{fig-proportion}Summary of the relationship between JSO
hyper-parameters on optimization success rate, using the holes index for
pipe data of dimensions 4-12. Bootstrap samples show the variability in
success rate. Success rate mostly plateaus by 50 jellyfish. The JSO has
some difficulty finding the pipe when dimension is higher than 8 and
maximum number of iteration has little effect.}

\end{figure}

The effect of JSO hyper-parameters (number of jellyfish and the maximum
number of iteration) on the success rate is summarized in
Figure~\ref{fig-proportion}. Bootstrap resampling is used to quantify
the uncertainty of the success rate through 500 bootstrap samples with
replacement from the original 50 simulation results for each task:
optimizing a high-dimensional data (d = 4, 6, 8, 10, or 12) with
different number of jellyfish ( 20, 50, or 100) and maximum number of
iteration (50 or 100). As the number of jellyfish and maximum iteration
allowed increase, the success rate also increases. For problems in lower
dimensional (4, 6) search spaces, small parameter values (20 jellyfish
and 50 iterations) are sufficient to achieve a high success rate. Many
more jellyfish (100 jellyfish and 100 iterations) are needed for finding
the pipe structure in d = 10, and 12. This suggests that even JSO will
struggle with high dimensions, and require substantially more
computational time to have an acceptable success rate, especially for
difficult PP indexes based on scagnostics.

\begin{table}

\caption{\label{tbl-smoothness-squintability}The columns \(\nu\) and
\(\varsigma\) show the smoothness and squintability measures for PP
indexes for the sine wave data. Only three are shown for illustration,
and a full table can be found in Supplementary materials. The splines
index has the largest squint angle and is the smoothest. The skinny
index has low smoothness and smallest squint angle.}

\centering{

\begin{tabular}{|llr|rr|llr|rr|llr|rr|llr|rr|llr|rr}
\toprule
shape & index & d & $\nu$ & $\varsigma$\\
\midrule
sine & splines & 4 & 3.5489 & 0.5723\\
sine & splines & 6 & 3.1149 & 0.5415\\
sine & splines & 8 & 2.7400 & 0.5394\\
\hline
sine & TIC & 4 & 3.4854 & 0.3839\\
sine & TIC & 6 & 3.0807 & 0.3747\\
sine & TIC & 8 & 2.7223 & 0.3785\\
\hline
sine & skinny & 4 & 2.5213 & 0.0810\\
sine & skinny & 6 & 2.2352 & 0.0973\\
\bottomrule
\end{tabular}

}

\end{table}%

\begin{figure}

\centering{

\pandocbounded{\includegraphics[keepaspectratio]{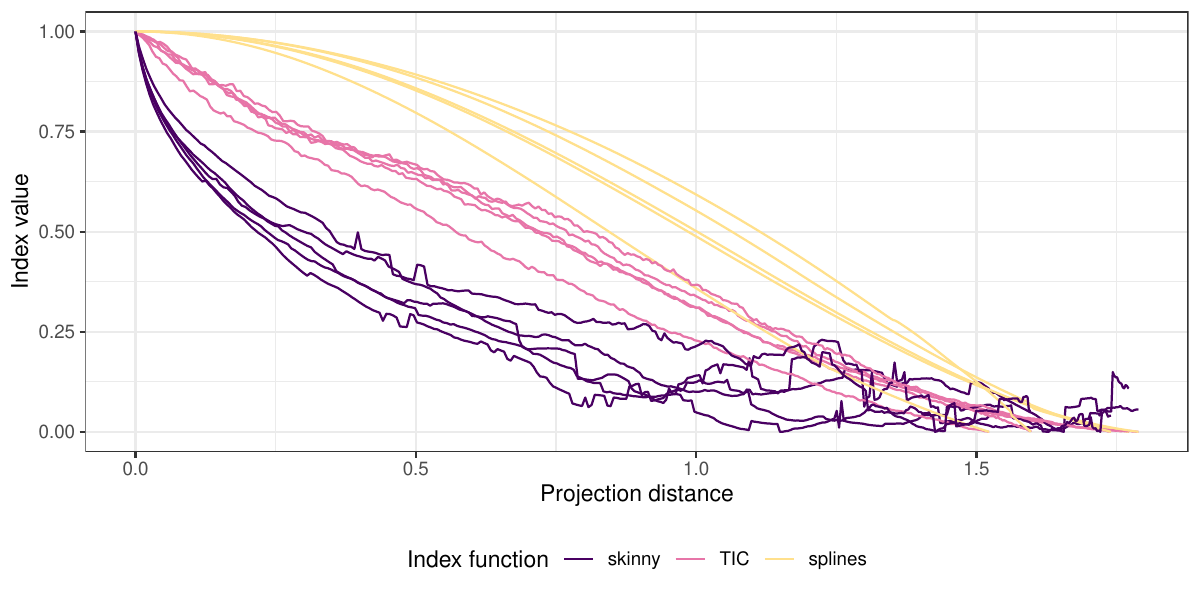}}

}

\caption{\label{fig-idx-proj-dist}Five traces of index value against
projection distance for each of the three indexes (TIC, spline, and
skinny) on 6D sine wave data. The traces of the splines index are
smooth, with a gradual change between noise and otimal projection. In
contrast, the skinny index has a noisy trace that increases rapidly near
the optimum. This illustrates the usefulness of the new measures for
squintability and smoothness.}

\end{figure}%

\begin{table}[!ht]

\caption{\label{tbl-mod-output}Jellyfish success rate relative to index
properties and jellyfish hyper-parameters. This is the summary from a
logistic regression fit to smoothness, squintability, dimension, number
of jellyfish and maximum number of iterations. Interestingly,
squintability and dimension strongly affect jellyfish optimization
success. The number of jellies marginally affects success, but index
smoothness, and increasing the number of iterations do not.}

\centering{

\fontsize{12.0pt}{14.4pt}\selectfont
\begin{tabular*}{\linewidth}{@{\extracolsep{\fill}}lcc}
\toprule
\textbf{Characteristic} & \textbf{OR} \textbf{(95\% CI)} & \textbf{p-value} \\ 
\midrule\addlinespace[2.5pt]
Smoothness & 1.01 (0.92 to 1.11) & 0.810 \\ 
Squintability & 1.23 (1.12 to 1.38) & {\bfseries <0.001} \\ 
Dimension & 0.58 (0.45 to 0.73) & {\bfseries <0.001} \\ 
Number of jellyfish & 1.32 (1.17 to 1.50) & {\bfseries <0.001} \\ 
Maximum number of iterations & 1.06 (0.94 to 1.20) & 0.328 \\ 
\bottomrule
\end{tabular*}
\begin{minipage}{\linewidth}
Abbreviations: CI = Confidence Interval, OR = Odds Ratio\\
\end{minipage}

}

\end{table}%

\begingroup\fontsize{10}{12}\selectfont

\begin{longtable}[t]{rrrr}

\caption{\label{tbl-joint-output}Fit statistics for the model.}

\tabularnewline

\\
\toprule
\textbf{Deviance} & \textbf{DF Residual} & \textbf{Null Deviance} & \textbf{DF Null}\\
\midrule
20.89 & 71 & 49.77 & 76\\
\bottomrule

\end{longtable}

\endgroup{}

\subsection{Effect of index properties on JSO success
rate}\label{effect-of-index-properties-on-jso-success-rate}

Table~\ref{tbl-smoothness-squintability} presents values computed for a
subset of PP indexes on the sine wave data for dimensions 4-8. (A full
table of the parameters estimated from the Gaussian process and from the
scaled logistic function for calculating smoothness and squintability
for all PP indexes and data combinations is included in the
supplementary materials). The column \(\nu\) is used as the smoothness
metric and the column \(\varsigma\) is calculated as equation
(\ref{eq-squintability-parametric}) as the squintability metric.

Figure~\ref{fig-idx-proj-dist} shows a sample of traces from the indexes
in Table~\ref{tbl-smoothness-squintability} for the 6D sine wave data.
This traces the PP index values computed for projections along an
interpolated path between a randomly generated noise projection and the
optimal projection (sine wave). (Note index values were
self-standardized to range between 0 and 1, for comparison.) The splines
index is very smooth with a gradual increase in value between the noise
projection and the optimal projection. The TIC index is slightly less
smooth and slightly less gradual change in value. Least desirable is the
skinny index which is noisier and the optimization needs to be much
closer to the optimal angle to have a higher value. This illustrates how
the new measures for squintability and smoothness work to describe the
PP indexes.

Table~\ref{tbl-mod-output} presents the results from fitting a logistic
regression model using the proportion of success as the response and
smoothness, squintability, dimension, number of jellyfish and maximum
number of iteration, as predictors. The fit suggests that the JSO
success rate is only affected by squintability, dimension and number of
jellyfish. As expected, the JSO success rate is higher with higher
squintability and/or more jellyfish, and lower when dimension is higher
(higher-dimensional optimization is more difficult). Interestingly, the
JSO is unaffected by the smoothness of the index function. This is
consistent with the way random search algorithms jump from value to
value without taking local regularity into account, as opposed to
gradient-based optimizers for example. Allowing for more iterations also
does not affect success rate significantly. A unit increase in
squintability increases the success rate by 23\%. As dimension increases
by one the success rate almost halves. Increasing the number of jellies
by 10 increases the success rate by 32\%. Table~\ref{tbl-joint-output}
shows the fit statistics supporting that the model is reasonably strong.

In relation to the model fit, it should be noted that there is moderate
correlation between squintability and smoothness, but this does not
substantially affect the fit. Fitting smoothness alone results in this
variable being weakly significant for explaining success rate, but the
effect is dominated by squintability. The addition of interaction terms
for squintability and data dimension, and smoothness and data dimension,
improves the model slightly, but not enough to be practically important
or change the interpretation above. The CRS optimizer couldn't be used
here because it is not efficient (Figure~\ref{fig-proj}) but we would
expect that smoothness does not overly affect it either. Both JSO and
CRS are expected to be robust to small local fluctuations.

These results also support the use of the proposed measures to
effectively capture key factors affecting the success rate of PP.

\section{Practicalities}\label{sec-discussion}

Using the JSO optimizer for PPGT in the \texttt{tourr} package requires
a change in user behavior. For all the existing optimizers a single
optimization path is followed. The JSO has multiple optimization paths,
and needs additional tools to incorporate into the PPGT, as explained
here.

\subsection{Using the JSO in a PPGT}\label{using-the-jso-in-a-ppgt}

To use JSO for PPGT in the \texttt{tourr} package, specify
\texttt{search\_f\ =\ search\_jellyfish} in the guided tour. The
animation function \texttt{animate\_*()} won't directly render the
animated projection for JSO, as with other existing optimizers since
multiple tour paths will need to be generated. To visualize the path
visited by each individual jellyfish, assign the animation to an object
(\texttt{res\ \textless{}-\ animate\_xy(...)}). This will save all the
bases visited by JSO, along with relevant metadata, as a tibble data
object (see Zhang et al. 2021 for more details on the data object).
Users can then view the path of individual jellyfish through extracting
the relevant bases recorded and construct it as a planned tour using the
\texttt{planned\_tour()} function (see the examples via
\texttt{?search\_jellyfish}). Users may wish to view the optimization
path of the most successful jellyfish, or a random selection of
jellyfish, or some specific runs that find local maxima.

\subsection{Computing the index properties for your new
index}\label{computing-the-index-properties-for-your-new-index}

The \texttt{ferrn} package (Zhang et al. 2021) has now included
functionality for computing the smoothness and squintability metrics.
Both metrics require sampling random bases using the
\texttt{sample\_bases()} function, followed by the calculation with
\texttt{calc\_smoothness()} or \texttt{calc\_squintability()}.

\subsubsection{Smoothness}\label{smoothness}

To sample bases for calculating smoothness, the following inputs are
required: the index function, the dataset, and the number of random
bases to sample. The output of \texttt{sample\_bases()} is a data frame
with a list-column of sampled bases and index values. Parallelization is
available to speed up the index value calculation through the
\texttt{parallel\ =\ TRUE} argument.

The \texttt{calc\_smoothness()} function takes the output from
\texttt{sample\_bases()} and fits a Gaussian process model to the index
values against the sampled bases, as the location, to obtain the
smoothness metric. Starting parameters and additional Gaussian process
arguments can be specified and details of the fit can be accessed
through the \texttt{fit\_res} attribute of the output.

\subsubsection{Squintability}\label{squintability}

Bases sampling for calculating squintability includes an additional step
of interpolating between the sampled bases and the optimal basis. This
step is performed when the arguments \texttt{step\_size} and
\texttt{min\_proj\_dist} in the \texttt{sample\_bases()} function are
set to non-NA numerical values. Given the projection distance typically
ranging from 0 to 2, it is recommended to set \texttt{step\_size} to 0.1
or lower, and \texttt{min\_proj\_dist} to be at least 0.5 to ensure a
meaningful interpolation length.

The \texttt{calc\_squintability()} function computes squintability using
two methods: 1) parametrically, by fitting a scaled sigmoid function
through non-linear least squares (\texttt{method\ =\ "nls"}), and 2)
non-parametrically, using kernel smoothing (\texttt{method\ =\ "ks"}). A
\texttt{bin\_width} argument is required to average the index values
over the projection distance before the fitting. For the parametric
case, the output provides the estimated parameters for the logistic
function (\(\theta_1\) to \(\theta_4\)) and the calculated squintability
metric as Equation (\ref{eq-squintability-parametric}). For the
non-parametric case, it shows the maximum gradient attained
(\texttt{max\_d}), the corresponding projection distance
(\texttt{max\_dist}), and the squintability metric as their products.

\section{Conclusion}\label{sec-conclusion}

This paper has presented new metrics to mathematically define desirable
features of PP indexes, squintability and smoothness, and used these to
assess the performance of the new JSO. The metrics will be generally
useful for characterizing PP indexes, and help with developing new
indexes.

In the comparison of the JSO against the currently used CRS, as expected
the JSO vastly outperforms CRS, and provides a high probability of
finding the global optimum. The JSO obtains the maximum more cleanly,
with a slightly higher index value, and plot of the projected data
showing the structure more clearly.

The JSO performance is affected by the hyper-parameters, with a higher
chance of reaching the global optimum when more jellyfish are used and
the maximum number of iteration is increased. However, it comes at a
computational cost, as expected. The performance declines if the
projection dimension increases and if the PP index has low
squintability. The higher the squintability the better chance the JSO
can find the optimum. However, interestingly smoothness does not affect
the JSO performance.

Future work can focus on developing more efficient optimization
algorithm for noise indexes with low squintability metric, which
resemble a needle-in-a-haystack problem (Siemenn et al. 2023).
Additionally, if the index function is differentiable, automatic
differentiation can be used to implement new gradient-based optimizers
(Prince 2023).

\section{Acknowledgement}\label{acknowledgement}

The article has been created using Quarto (Allaire et al. 2022) in R (R
Core Team 2023). The source code for reproducing the work reported in
this paper can be found at:
\url{https://github.com/huizezhang-sherry/paper-jso}. The simulation
data produced in Section 5 can be found at
\url{https://figshare.com/articles/dataset/Simulated_raw_data/26039506}.
Nicolas Langrené acknowledges the partial support of the Guangdong
Provincial/Zhuhai Key Laboratory IRADS (2022B1212010006) and the BNBU
Start-up Research Fund UICR0700041-22.

The R packages used in this work include: \texttt{tidyr} (Wickham,
Vaughan, and Girlich 2024), \texttt{dplyr} (Wickham et al. 2023),
\texttt{ggplot2} (Wickham 2016), \texttt{knitr} (Xie 2014),
\texttt{gtsummary} (Sjoberg et al. 2021), \texttt{patchwork} (Pedersen
2024), \texttt{ggh4x} (Van Den Brand 2024), \texttt{broom} (Robinson,
Hayes, and Couch 2024), \texttt{kableExtra} (Zhu 2024), \texttt{ferrn}
(Zhang et al. 2021), and \texttt{cassowaryr} (Mason et al. 2022).

\section*{Supplementary materials}\label{supplementary-materials}
\addcontentsline{toc}{section}{Supplementary materials}

The supplementary materials available at
\url{https://github.com/huizezhang-sherry/paper-jso} include: 1) details
of the indexes used in the simulation study, 2) the table of the
Gaussian process parameters and logistic function parameters for all 23
PP problems investigated, 3) the script to get started with using JSO in
a PPGT and calculating smoothness and squintability, as explained in
Section~\ref{sec-discussion}, along with 4) the full code to reproduce
the plots and summaries in this paper.

\section*{References}\label{references}
\addcontentsline{toc}{section}{References}

\phantomsection\label{refs}

\end{document}